\documentclass[reprint,amsmath,amssymb,aps,prm]{revtex4-1}

\usepackage{graphicx}
\usepackage{dcolumn}
\usepackage{scrextend}
\usepackage[colorlinks=true, citecolor=blue, linkcolor=red]{hyperref}
\usepackage{bm}
\setcitestyle{numbers,square}
\usepackage{longtable}

\newcommand{\bj}[1]{\textcolor{blue}{#1}}

\begin{document}
\title{High Flexoelectric Constants in Janus Transition-Metal Dichalcogenides}
\author{Brahmanandam Javvaji}
\thanks{Brahmanandam Javvaji and Bo He contributed equally to this work.}
\author{Bo He} 
\thanks{Brahmanandam Javvaji and Bo He contributed equally to this work.}
\affiliation{Institute of Continuum Mechanics, Leibniz Universit\"at Hannover, Appelstr. 11, 30167 Hannover, Germany}
\author{Xiaoying Zhuang}
\email{Corresponding author: zhuang@ikm.uni-hannover.de}
\affiliation{Institute of Continuum Mechanics, Leibniz Universit\"at Hannover, Appelstr. 11, 30167 Hannover, Germany}
\affiliation{College of Civil Engineering, Tongji University, 1239 Siping Road, 200092 Shanghai, China}
\author{Harold S. Park}
\email{Corresponding author: parkhs@bu.edu}
\affiliation{Department of Mechanical Engineering, Boston University, Boston, Massachusetts 02215, USA}

\begin{abstract}  
\noindent Due to their combination of mechanical stiffness and flexibility, two-dimensional (2D) materials have received significant interest as potential electromechanical materials.  Flexoelectricity is an electromechanical coupling between strain gradient and polarization. Unlike piezoelectricity, which exists only in non-centrosymmetric materials, flexoelectricity theoretically exists in all dielectric materials.   However, most work on the electromechanical energy conversion potential of 2D materials has focused on their piezoelectric, and not flexoelectric behavior and properties.  In the present work, we demonstrate that the intrinsic structural asymmetry present in monolayer Janus transition metal dichalcogenides (TMDCs) enables significant flexoelectric properties.  We report these flexoelectric properties using a recently developed charge-dipole model that couples with classical molecular dynamics simulations.  By employing a prescribed bending deformation, we directly calculate the flexoelectric constants while eliminating the piezoelectric contribution to the polarization.  We find that the flexoelectric response of a Janus TMDC is positively correlated to its initial degree of asymmetry, which contributes to stronger $\sigma-\sigma$ interactions as the initial degree of asymmetry rises. In addition, the high transfer of charge across atoms in Janus TMDCs leads to larger electric fields due to $\pi-\sigma$ coupling.  These enhanced $\sigma-\sigma$ and $\pi-\sigma$ interactions are found to cause the flexoelectric coefficients of the Janus TMDCs to be several times higher than traditional TMDCs such as MoS$_{2}$, whose flexoelectric constant is already ten times larger than graphene.   

\end{abstract}


\maketitle


\section{Introduction}
\label{sec:introduction}

Atomically thin two-dimensional (2D) materials have a variety of unique physical properties that have made them attractive for many different applications \cite{pop2012thermal,ghosh2010dimensional,fei2014strain,yu20172d,wang2012electronics,hanakata2016polarization,andrew2012mechanical,javvaji2016mechanical,kim2015materials,sun2016optical,mas20112d}.  An emerging area of interest for 2D materials is electromechanical coupling, due to the desire to miniaturize sensors and actuators to the micro and nanoscales.  The most widely studied electromechanical coupling mechanism is piezoelectricity, which has also been investigated for 2D materials, including graphene, hexagonal boron nitride (h-BN), transition metal dichalcogenides (TMDCs) and many others \cite{Wang2015observattion,Song2017piezo,Zheng2017hexagonal,Zelisko2014anomalous,blonsky2015ab,Kundalwal2017,chandratre2012coaxing,duerloo2013flexural,zhang2017boron,zhou2016theoretical,javvaji2018generation,wu2014piezoelectricity,duerloo2012intrinsic}.  We note that the majority of this work has focused on the in-plane electromechanical properties.

TMDCs exhibit a unique three-layer atomic arrangement, where a metallic (M) atom symmetrically bonds with two chalcogenide atoms (X and X) in the out-of-plane direction.  This structure makes them as good candidates for in-plane stretching based piezoelectric materials \cite{blonsky2015ab}, though a recent experimental study measured how an out-of-plane indentation induced an electrical response in MoS$_2$ \cite{Brennan2017out}. This deformation induced significant changes in the bond lengths between M and X atoms, resulting in a strain gradient, which potentially enabled flexoelectricity, which is a form of electromechanical coupling in which electrical polarization is generated due to strain gradients \cite{majdoub2008enhanced,He2018,javvaji2018generation,Kalinin2008,dumitricua2002curvature,kundalwal2017strain}. Recent work from the present authors \cite{zhuang2019intrinsic} investigated the bending flexoelectricity in various 2D materials, including graphene, graphene allotropes, nitrides, graphene analogs of group-IV elements and TMDCs.  That study found that MoS$_2$ (an MXX material) has a flexoelectric coefficient that is ten times larger than graphene due to enhanced charge transfer resulting from asymmetrical bending-induced changes in the M-X bond lengths. 

Because of the impact of asymmetry in the bending-induced changes in the M-X bond lengths in enhancing the flexoelectric properties of MoS$_{2}$, we focus here on potential flexoelectric effects in another class of TMDCs, the so-called Janus TMDCs, which introduce an asymmetry in the MXX by replacing one of the X layers of atoms with a different chalcogenide atom Y, resulting in an MXY structure and intriguing physical properties \cite{Wang2018,Guo2018,Jin2018,Shi2018,Huang2018}, as recently reviewed \cite{Li2018}.  The changes in the atomic mass and electronic configurations of X and Y atoms in MXY generates an out-of-plane as well as in-plane dipole moment, which are absent in the MXX structure, due to the non-cancellation of interactions between M-X and M-Y. Density functional theory (DFT) simulations have been used to report high piezoelectric coefficients for various materials in the Janus TMDCs family compared to conventional TMDCs \cite{dong2017large}, while a recent experimental report reported the out-of-plane piezoelectric response for a Janus MoSSe monolayer \cite{Lu2017}. Furthermore, multi-layer Janus TMDCs have shown very high out-of-plane piezoelectric coefficients due to the increase in vertical dipole moments \cite{dong2017large}. Overall, an electromechanical imbalance exists due to the element changes between X and Y in MXY, which should facilitate asymmetric deformation, strain gradients, and thus flexoelectricity, while also impacting the induced dipole moments in the Janus TMDCs.

Various works have investigated the piezoelectric properties of Janus TMDCs.  For example, the tensile load induced in-plane piezoelectric coefficient ($d_{11}$) of the Janus TMDCs was found to be several times larger than their out-of-plane shear piezoelectric coefficient ($d_{31}$) \cite{dong2017large,zhang2019first}. where the $d_{31}$ coefficient is absent in most other 2D materials, including  MoS$_2$, due to reflection symmetry\cite{blonsky2015ab}. Additionally, the out-of-plane piezoelectric coefficient ($d_{33}$) for monolayer MoSSe under compression was reported to be several orders of magnitude smaller than $d_{11}$ \cite{Lu2017}. Under compression, large values for $d_{33}$ were reported for multi-layer Janus TMDCs, where the monolayers are arranged such that the induced polarization does not cancel out. However, for the multilayer structure, the out-of-plane elastic constants increase with increasing number of layers, which requires an increasing amount of force to induce deformation in the vertical direction \cite{hu2017peculiar}, and thus is not an optimal choice for electromechanical energy conversion. In contrast, Janus TMDCs, as with most atomically thin 2D materials, are significantly easier to bend rather than stretch\cite{dong2017large,Jiang2013,Brennan2017out}.  The relative ease of bending 2D materials, coupled with the fact that bending intrinsically generates a strain gradient, indicates significant electromechanical energy conversion potential for bending flexoelectricity as compared to in-plane piezoelectricity for Janus TMDCs.

In order to compare the energy conversion potential between flexoelectricity and piezoelectricity, it is essential to separate the contribution of the flexoelectric response.  However, the effect of flexoelectricity is measured in terms of effective piezoelectricity in practice. For example, the experimental work \cite{Brennan2017out} determined the flexoelectric coefficient through a relationship with the measured piezoelectric coefficient under assumptions of small length scales and linear electric fields. Additional example \cite{Jin2018virtual} reported the effective out-of-plane piezoelectricity from MoS$_2$. The experimental study reported the piezoelectric responses of a corrugated TMDC \cite{Kang2018,Kang2019}, where electrical polarization is mainly due to the local strain gradients that govern flexoelectricity. A recent study addressed the effective piezoelectricity from the effect of flexoelectricity in a non-piezoelectric material \cite{Abdollahi2019}. Large-scale experimental approaches like axial stretching or radial compression of a cylindrical rod wrapped with a non-piezoelectric material could enable direct measurements of the  flexoelectric coefficients \cite{Zhang2017}. However, experimental methods to find the flexoelectric constants at the nanoscale are unresolved. First-principle simulations can isolate the flexoelectric effect by assuming unstable wrinkles in TMDCs \cite{Shi2018flexo}. The authors previous work \cite{zhuang2019intrinsic} provides a mechanical bending deformation which enables calculation of the flexoelectric response by removing the out-of-plane piezoelectric contribution to the total polarization. 

In this work, we coupled classical molecular dynamics (MD) with a charge-dipole model \cite{mayer2006charge,mayer2007formulation} to investigate the bending flexoelectric response of the Janus TMDC family.  We first validate the simulation methodology with respect to previous, DFT-calculated in-plane piezoelectric coefficients \cite{dong2017large}.  We then propose a mechanical bending deformation to enable the direct measurement of the flexoelectric response by eliminating the piezoelectric contribution to the polarization.  Our results show that the bending flexoelectric constants of Janus TMDCs are significantly higher than that of traditional TMDCs such as MoS$_{2}$. The flexoelectric enhancement is found to emerge from the structural asymmetry that is intrinsic to Janus TMDCs, which results in both stronger $\sigma-\sigma$ and $\pi-\sigma$ interactions than is found in traditional TMDCs. 

\section{Simulation method}
\label{sec:simulation_method}

We introduce in this section the computational model we use to calculate the electrical polarization due to mechanical deformation.  Specifically, a combination of short-range bonded interactions with long-range charge-dipole (CD) interactions were considered for calculating the forces acting on a given atomic system. According to the CD model \cite{mayer2006charge,mayer2007formulation}, each atom is assumed to carry a charge $q$ and dipole moment $\mathbf{p}$.   The short-range atomic interactions are modeled using a Stillinger-Weber potential \cite{jiang2019misfit}, which was previously used to study the misfit strain induced buckling for lateral heterostructures with different combinations of TMDCs (MoS$_2$-WSe$_2$; MoS$_2$-WTe$_2$; MoS$_2$-MoSe$_2$ and MoS$_2$-MoTe$_2$).  This potential was shown to capture the previously reported spontaneous curling behavior of Janus TMDCs~\cite{xiong2018spontaneous}.  The CD model requires a parameter $R$ (related to atomic polarizability $\alpha$) to evaluate the charge and dipole for each atom, which is obtained by matching the calculated polarizability $(\alpha_{\text{Cal.}})$ with DFT calculated value $(\alpha_{\text{DFT}})$, where the complete details about the process for calculating the CD parameters are given in Ref.~\cite{zhuang2019intrinsic}. Table \ref{table:atomic_polar} lists the lattice parameters of each Janus TMDC,  the parameter $R$, $\alpha_{\text{DFT}}$ and $\alpha_{\text{Cal.}}$. All the simulations in this study were conducted in the open-source molecular dynamics code LAMMPS \cite{Plimpton1995}. The GAUSSIAN software \cite{frisch2016gaussian} was employed to estimate $\alpha_{\text{DFT}}$. The details of the atomic forces resulting from the CD model can be found in \cite{javvaji2018generation} and references therein. \bj{The current DFT calculations do not consider spin-orbit coupling (SOC). We tabulated the spontaneous dipole moment for Janus TMDCs from the present CD model $(p^0_{\text{Cal.}})$ and recent DFT reports with and without SOC $(p^0_{\text{DFT}})$. While SOC does have some effect on the results, Table.~\ref{table:atomic_polar} shows that the comparison between $p^0_{\text{Cal.}}$ and $p^0_{\text{DFT}}$ is quite satisfactory. This shows that the present CD model is sufficiently accurate to predict the flexoelectric properties.}

\begin{table*}
	\begin{ruledtabular}
	\centering
	\caption{\bj{The spontaneous dipole moment from the present CD model $(p^0_{\text{Cal.}}~\text{in e\AA})$ and recent DFT reports with and without SOC $(p^0_{\text{DFT}}~\text{in e\AA})$.} Total polarizability estimated by DFT ($\alpha_{\text{DFT}}$ in $\text{\AA}^3$) and the present CD model ($\alpha_{\text{Cal.}}$ in $\text{\AA}^3$)~\cite{zhuang2019intrinsic}. Atomic polarizability ($R_i$ in \AA) and lattice parameters (a,b are the lattice constants, $l_1$ and $l_2$ are the bond lengths for M-X and M-Y in $\text{\AA}$, respectively) for the Janus TMDCs. For simplicity, in this study the Janus TMDCs are denoted as MXY with M = Mo and W; X,Y = S,Se and Te; X represents the chalcogenide atom with smaller atomic number while Y represents the chalcogenide atom with larger atomic number.}
	\label{table:atomic_polar}
	\begin{tabular}{c c c c c c c c c c c c }
		Material & \bj{$p^0_{\text{Cal.}}$} & \bj{$p^0_{\text{DFT}}$} & $\alpha_{\text{DFT}} $ & $\alpha_{\text{Cal.}} $ & $R_M$  & $R_X$  & $R_Y$ & a,b & $l_1$ & $l_2$  \\
		\hline
		MoS$_2$ & 0.0 & 0.0\footnote{\label{a}Reference with SOC \cite{Defo2016}} & 12.320 & 12.350 & 0.69 & 1.04 & 1.04 & 3.160\footnote{\label{b}Reference \cite{hu2018intrinsic}} & 2.420\footref{b} & 2.420\footref{b} \\
		\hline
		MoSSe  & 0.032 & 0.0391\footnote{\label{c}Reference with SOC \cite{Xia2018}}$^,$\footnote{\label{d}Reference without SOC \cite{Jin2018}}, 0.052\footnote{\label{e}Reference with SOC \cite{Li2017jpcs}}$^,$\footnote{\label{f}Reference without SOC \cite{Ji2018jpcc}} & 13.456 & 13.450 & 0.84 & 1.14 & 1.06 & 3.288\footref{b} & 2.416\footref{b} & 2.530\footref{b}  \\
		MoSTe  &0.0432 & 0.0412\footref{c} & 15.590 & 15.593 & 1.0  & 1.16 & 1.02 & 3.343\footref{b} & 2.432\footref{b} & 2.715\footref{b}  \\
		MoSeTe &0.0649 & 0.079\footref{c} &16.768 & 16.769 & 1.04 & 1.12 & 1.05 & 3.412\footref{b} & 2.552\footref{b} & 2.717\footref{b}  \\
		\hline
		WSSe   &0.0479 & 0.0362\footref{c}, 0.05\footref{e} & 16.824 & 16.822 & 0.94 & 1.18 & 1.08 & 3.232\footref{b} & 2.421\footref{b} & 2.538\footref{b}  \\
		WSTe   &0.0574 & 0.0389\footref{c} & 19.489 & 19.489 & 1.08 & 1.18 & 1.08 & 3.344\footref{b} & 2.438\footref{b} & 2.720\footref{b}  \\
		WSeTe  &0.0587 & 0.0757\footref{c} & 21.225 & 21.224 & 1.08 & 1.24 & 1.08 & 3.413\footref{b} & 2.559\footref{b} & 2.722\footref{b}  
	\end{tabular}	
	\end{ruledtabular}	
\end{table*}

\begin{figure*}
	\centering
	\includegraphics[width=1.0\linewidth]{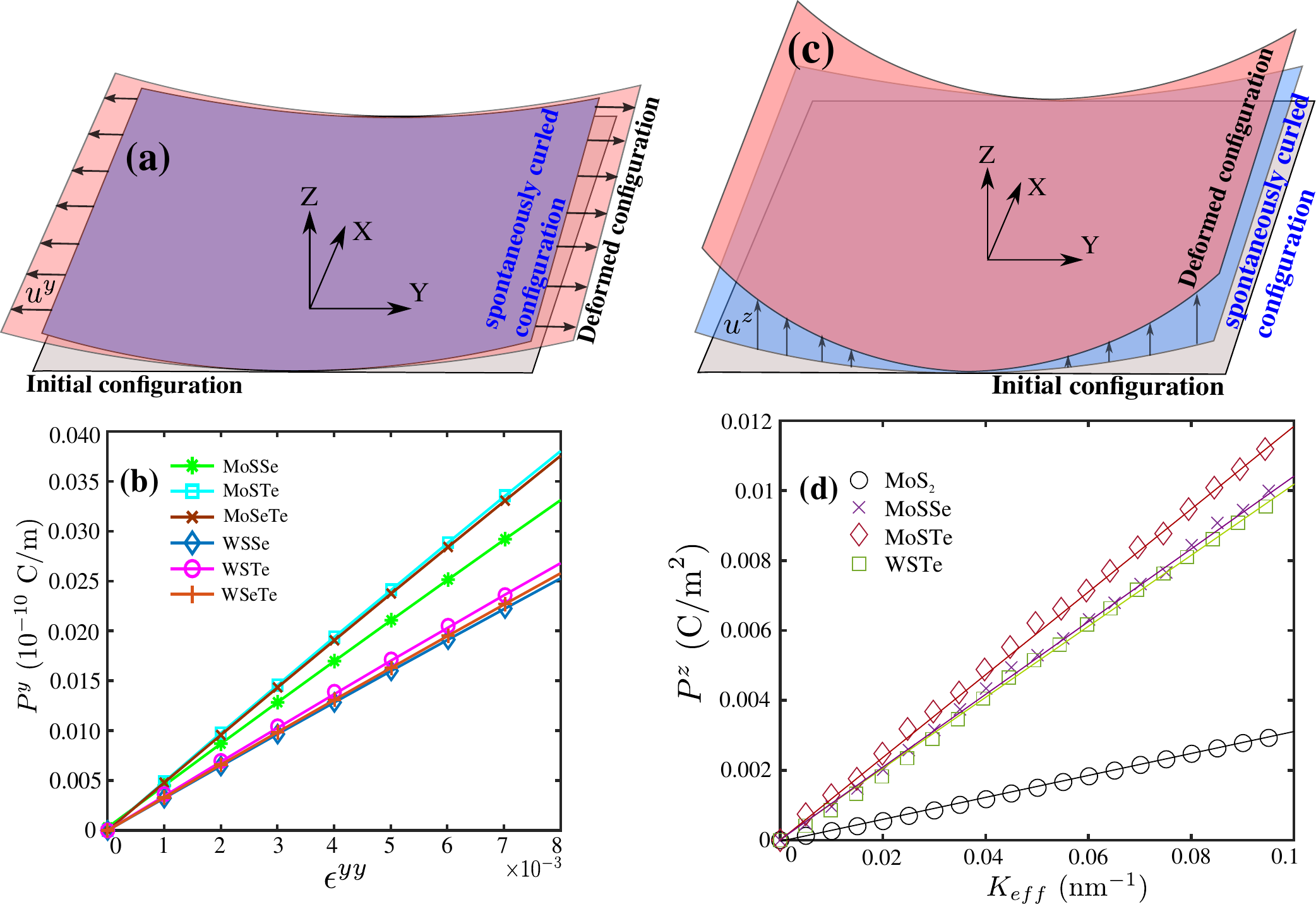}
	\caption{(a) Loading scheme for calculating the in-plane piezoelectric coefficient $d_{11}$ and (b) polarization $P^y$ (C/m) vs strain $\epsilon^{yy}$ for the Janus TMDC systems. (c) Schematic of applied bending deformation to calculate the flexoelectric constants. (d) Polarization $P^z$ (C/m$^2$) versus strain gradient $K_{eff}~(\text{nm}^{-1})$ for MoS$_2$, MoSSe, MoSTe and WSTe. Markers indicate the data from CD model and soild lines indicate the linear fitting.}
	\label{fig:piezo-flexo}
\end{figure*}

We first validated the CD model by comparing to previously reported in-plane piezo coefficients \cite{dong2017large}. In this study, the in-plane piezoelectric coefficients of the Janus TMDCs (MXY, where M = Mo, W; X, Y = S, Se, and Te, where the atomic mass of X is smaller than the atomic mass of Y) are obtained by subjecting a square Janus TMDC sample with dimensions 80 \AA $\times$ 80 \AA~to tensile loading in the y-direction as shown in Fig.~\ref{fig:piezo-flexo}(a). The initial configuration is flat, and the relaxed atomic configurations of the Janus TMDC are obtained through energy minimization, after which the atomic charge $q_i$ and dipole moments $\mathbf{p}_i$ are derived from the CD model for atom $i$.  The relaxed configuration is bent due to spontaneous curling~\cite{xiong2018spontaneous, Wang2018}, which arises from the structural asymmetry between the M-X and M-Y layers, resulting in stretching in the Y layer and compression in the X layer. 

The total polarization $\mathbf{P}$ of the system is calculated by $\mathbf{P} = \frac{1}{V} \left(\sum_{i=1}^{n} \mathbf{p}_i\right)$ where $V$ is the volume and $n$ is the total number of atoms in the system. The in-plane strain is calculated by $\epsilon^{yy} = \frac{ly-ly_{\text{ini}}}{ly_{\text{ini}}}$, where $ly$ and $ly_{\text{ini}}$ are the deformed length and initial length in the y-direction, respectively. The loading scheme for calculating the in-plane piezoelectric coefficient and polarization ($P^y$) - strain ($\epsilon^{yy}$) diagrams are presented in Figs.~\ref{fig:piezo-flexo}(a) and (b). The polarization ($P^y$) - strain ($\epsilon^{yy}$) diagrams are shifted to have zero initial polarization by subtracting the polarization caused by the initial spontaneous curling \cite{xiong2018spontaneous}. A linear relation is observed between the polarization $P^y$ and given strain $\epsilon^{yy}$ in Fig.~\ref{fig:piezo-flexo}(b). The slope of the linear relation yields the in-plane piezoelectric coefficient of the Janus TMDCs. The calculated piezoelectric coefficients $(d_{yyy}~\text{or}~d_{11})$ are in good agreement with the reported DFT values \cite{dong2017large} (see Table \ref{table:piezo-flexo}), which validates the effectiveness of the CD model. 

\begin{table}
	\begin{ruledtabular}
		\centering
		\caption{In-plane piezoelectric coefficients $d_{yyy}$ ( $\times 10^{-10}$ C/m) for Janus TMDCs using proposed CD model in comparison to previous DFT results ($d_{yyy}^{\text{DFT}}$).  $\mu_{zyzy}$ (nC/m) are the bending flexoelectric constants while $l_2-l_1 \:(\text{\AA})$ is the initial asymmetry for Janus TMDCs.}
		\label{table:piezo-flexo}
			\begin{tabular}{c c c c c}
				Material & $d_{yyy}$ & $d_{yyy}^{\text{DFT}}$ & $l_2-l_1 $ &  $\mu_{zyzy}$ \\
				\hline
				MoS$_2$ & 3.95 & 3.56\footnote{\label{a}Reference \cite{dong2017large}} & 0.0 & 0.032 \\
				\hline
				MoSSe   & 4.099 & 3.74\footref{a} & 0.114 & 0.117 \\
				MoSeTe  & 4.676 & 4.35\footref{a} & 0.165 & 0.120 \\
				MoSTe   & 4.733 & 4.53\footref{a} & 0.283 & 0.125 \\
				\hline 
				WSSe    & 3.144 & 2.57\footref{a} & 0.117 & 0.089 \\
				WSeTe   & 3.209 & 3.34\footref{a} & 0.163 & 0.092 \\
				WSTe    & 3.327 & 3.48\footref{a} & 0.282 & 0.114 \\			
			\end{tabular}
	\end{ruledtabular}				
\end{table}

\section{Results and discussion}
\label{sec:result_discussion}

In this section, we describe the bending scheme used to study the flexoelectric properties of Janus TMDCs. To determine the bending flexoelectric coefficients for the Janus TMDCs, the loading scheme illustrated in Fig.~\ref{fig:piezo-flexo}(c) is applied.  Specifically, the following displacement field is applied to the atomic system
\begin{equation}
u^z = K\frac{y^2}{2},
\label{eq:out-bending}
\end{equation}
where $y$  represents the atom coordinate in the $y$ direction, and $\frac{1}{2} K$ represents the given strain gradient of the bending plane. We note that the imposed displacement field in Eq. (\ref{eq:out-bending}) is imposed starting from the relaxed, or spontaneously curved configuration in Fig.~\ref{fig:piezo-flexo}(c).  Once the deformation is prescribed, the boundary region atoms are held fixed while the interior atoms are allowed to relax to energy minimizing positions using the conjugate-gradient algorithm, after which the point charges $q_i$ and dipole moments $\mathbf{p}_i$ are found for each atom by CD model.  The bending flexoelectric constant $\mu_{zyzy}$ can then be obtained as the slope of the resulting plot between polarization and strain gradient.

\begin{figure}[h]
	\centering
	\includegraphics[width= 1.0\linewidth]{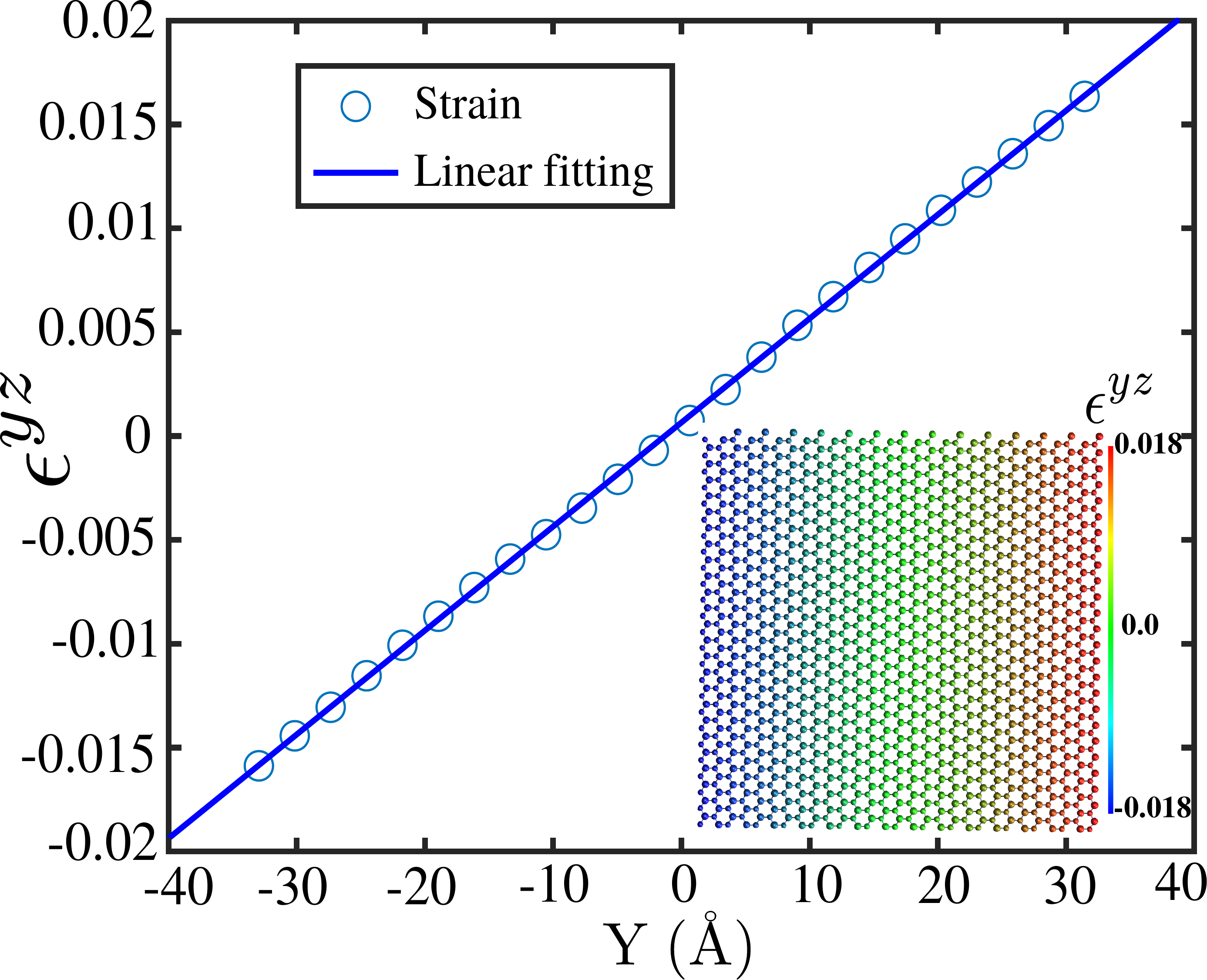}
	\caption{Strain profile $\epsilon_{yz}$ versus y coordinates $Y$.  Inset shows a contour plot of the strain $\epsilon_{yz}$.}
	\label{fig:strain_profile}
\end{figure}

In general, an imposed deformation would induce polarization, which has contributions from both piezoelectric and flexoelectric effects.  The current bending deformation in Eq.~(\ref{eq:out-bending}) induces only the strain component $\epsilon^{yz}$ and the strain gradient term $(\frac{\partial \epsilon^{yz}}{\partial y})$, while the remaining components of the strain and strain gradient tensors are zero. Therefore, the total induced polarization in the z-direction is 
\begin{equation}\label{eq:polar}
P^z = d_{zyz} \epsilon^{yz} + \mu_{zyzy} \frac{\partial \epsilon^{yz}}{\partial y},
\end{equation}	
where $d_{zyz}$ and  $\mu_{zyzy}$ are the bending piezoelectric coefficient and flexoelectric coefficient, respectively. The inset of Fig.~\ref{fig:strain_profile} represents the atomic configuration of MoSSe obtained using the OVITO software \cite{Stukowski_2009} when the applied curvature $K=0.05~\text{nm}^{-1}$.  The atoms are colored according to the $\epsilon^{yz}$ component of strain, which is calculated from Eq.~(\ref{eq:astrain}) as \cite{Falk1997,Shimizu2007}
\begin{equation}
\epsilon_i^{\alpha \beta} = \frac{1}{2} \left[ F_i^{\beta \alpha} F_i^{\alpha \beta} - \delta^{\alpha \beta}\right],
\label{eq:astrain}
\end{equation}
where $\epsilon_i$ is the atomic strain for atom $i$, $F_i^{\alpha \beta}$ is the $i$th atomic component of deformation gradient, $\delta$ is the Kronecker delta, $\alpha$ and $\beta$ are the coordinate components. 

As shown in Fig.~\ref{fig:strain_profile}, the strain $\epsilon^{yz}$ varies from -0.018 to 0.018 along the $y-$direction in the MoSSe sheet. These strain values represent averaged values of the atomic strain, which were found by averaging over 24 equal-width bins along the y-direction. The linear variation of atomic strain indicates that total strain is zero and maintains the point group symmetry posed by MXY \cite{dong2017large}. Overall, the observed symmetry removes the piezoelectric part of polarization $d_{zyz} \epsilon^{yz} = 0$, which supports the assumption that the imposed bending deformation removes the piezoelectric contribution to the polarization.

The slope of the plot in Fig. \ref{fig:strain_profile} between $\epsilon^{yz}$ and the $y-$coordinate of each bin leads to the effective strain gradient $K_{eff}=\partial \epsilon^{yz}/\partial y$. The current value of $K_{eff}$ differs from $\frac{1}{2}K$ (from Eq.~(\ref{eq:out-bending})) by about 15\%. Theoretically, the numerical value of $K_{eff}$ should be equal to $\frac{1}{2}K$ under the imposed bending deformation. For example, previous works found that for conventional TMDCs (MoS$_2$, WS$_2$ and CrS$_2$), the effective strain gradient is equal to half of the given value of $K$ \cite{zhuang2019intrinsic} under the same bending deformation. However, the observed difference in Janus TMDCs is due to the spontaneous curling effect \cite{xiong2018spontaneous, Wang2018}, which arises from the structural asymmetry between the M-X and M-Y layers, resulting in stretching in the Y layer and compression in the X layer. In order to account for the spontaneous deformation, the effective strain gradient $K_{eff}$ is used in our calculations of the flexoelectric coefficient. 

We note that a similar type of curling, i.e. ripplocations, have been observed in MoS$_2$ by introducing line defects (the removal of sulfur atoms along a line) \cite{Fabbri2016}, which generate local strain gradients and thus flexoelectricity. However, in the present work we restrict ourselves to calculations of intrinsic flexoelectricity for spontaneously curled Janus TMDCs.

\subsection{Flexoelectric effect in MoS$_2$ and MoSSe}
\label{sec:mos2vsmosse}

In this section, we examine MoS$_2$ and MoSSe to illustrate the effects of structural asymmetry on the resulting flexoelectric properties. MoS$_{2}$ is chosen as a TMDC that does not have structural asymmetry (MXX), while MoSSe is chosen as a representative Janus TMDC which does have structural asymmetry (MXY). To aid in the analysis of the resulting flexoelectric constants, we note that the dipole moment of an atom $\mathbf{p}$ depends primarily on three factors: the effective atomic polarizability ($R$), the charge induced electric field $(E^{q-z})$, and the dipole induced electric field $(E^{p-z})$. We focus, as shown previously in Eq. (\ref{eq:polar}), on the the out-of-plane (z-direction) dipole moment $p^z$ and associated polarization $P^z$.  

\begin{figure}[h]
	\centering
	\includegraphics[width= 1.0\linewidth]{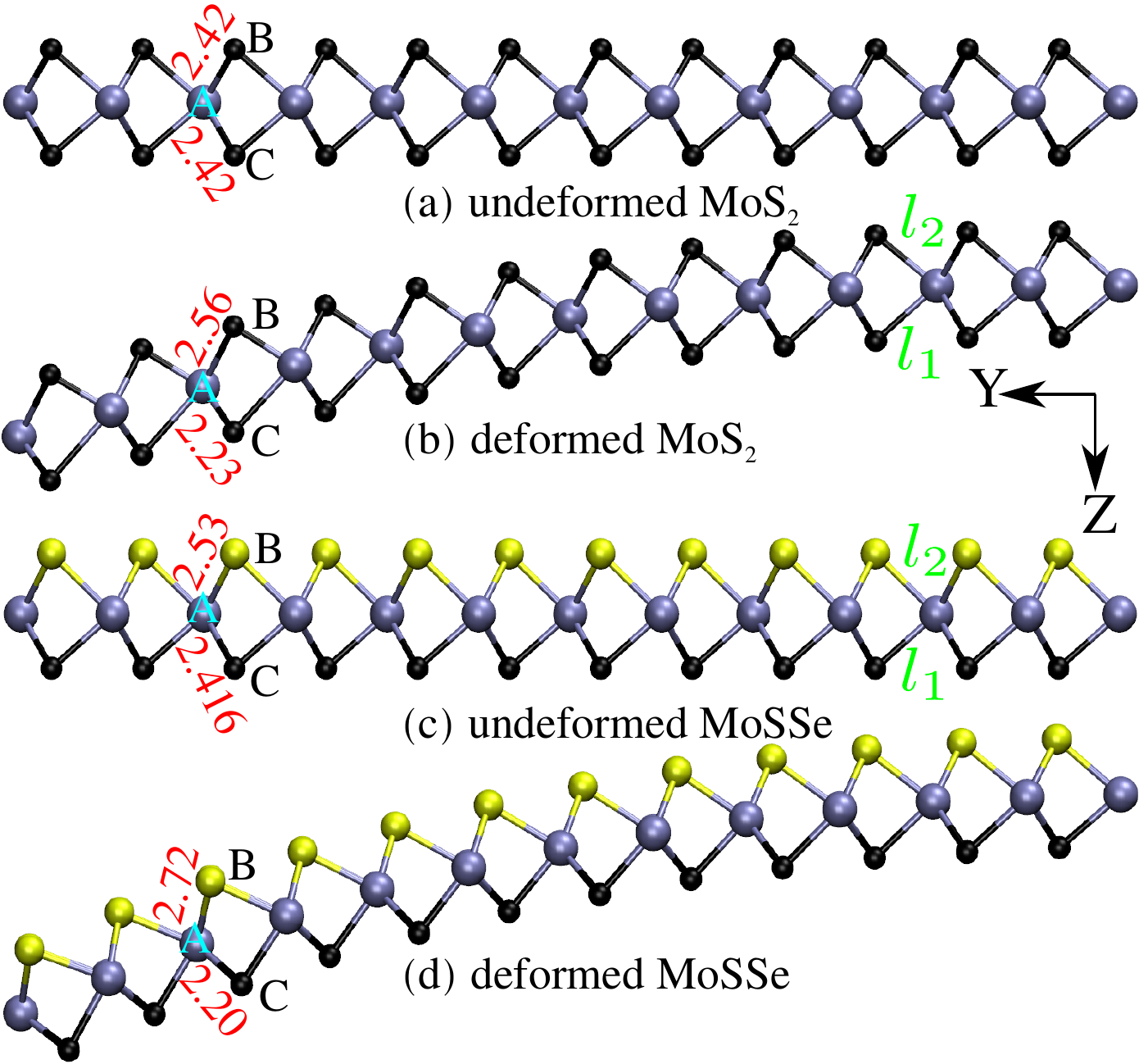}
	\caption{Atomic configurations of (a) undeformed MoS$_{2}$, (b) deformed MoS$_{2}$ at $K_{eff}$ = 0.05 nm$^{-1}$, (c) undeformed MoSSe and (d) deformed MoSSe at $K_{eff}$ = 0.05 nm$^{-1}$.}
	\label{fig:mos2_mosse}
\end{figure}

The bond length between atoms A-B and A-C in Fig.~\ref{fig:mos2_mosse}(a) is equal to $2.42~\text{\AA} $ $(l_{Mo-S})$ for unbent MoS$_{2}$.  Because the A-B and A-C bond lengths are the same, there is no initial structural asymmetry for MoS$_{2}$, and the bonds induce equal and opposite electric fields, which cause the total dipole moment, and thus total polarization of each MoS$_{2}$ unit cell to be zero.  However, significant changes in the bond lengths are observed in the deformed state (Fig.~\ref{fig:mos2_mosse}(b)) for a curvature of $K_{eff}$ = 0.05 nm$^{-1}$.  Bond A-B is stretched from 2.42 to 2.56~\text{\AA} while bond A-C is compressed from 2.42 to 2.23~\text{\AA}. This difference in bond lengths breaks the electric field symmetry and increases the total dipole moment.   At $K_{eff}$ = 0.05 nm$^{-1}$, the total  electric field difference $\Delta(E^{p-z}+E^{q-z})$ with respect to the initial (undeformed) configuration is 53.06 $\text{V/\AA}$, which increases the total polarization from $0$ to 0.0016 C/m$^2$. The changes to $E^{q-z}$ and $E^{p-z}$ at $K_{eff}$ = 0.05 nm$^{-1}$ are 45.6 and 7.4 $\text{V/\AA}$, respectively. The significant contribution of $E^{q-z}$ implies the increasing importance of $\pi-\sigma$ interactions in generating the dipole moment $p^z$. The $\pi-\sigma$ interactions originate from the coupling between valence electrons and bonding electrons \cite{Gleiter1987}, which are also interpreted as pyramidalization \cite{Surya2012,dumitricua2002curvature,Nikiforov2014}.  Furthermore, a recent DFT study reported \cite{Pike2017} electron transfer from the $p$ orbitals of S atoms to the $d_{z^2}$ orbital of Mo atoms.  This electron transfer modifies the charges on the Mo and S atoms, which generates local charge induced electric fields, which are captured through $E^{q-z}$ within the current CD model. Further changes in the bond length between Mo and S atoms enhance the $\pi-\sigma$ coupling \cite{Huang2018}, resulting in large $E^{q-z}$ and $P^z$. The variation of total polarization with the effective strain gradient for MoS$_2$ is given in Fig.~\ref{fig:piezo-flexo}(d), where the slope represents the flexoelectric constant for MoS$_{2}$ $(\mu_{\text{MoS}_2})$, which is found to be $0.032$ nC/m. 

In contrast to MoS$_{2}$, the Janus TMDC MoSSe has a structural asymmetry in the undeformed configuration between the Mo-S and Mo-Se atomic layers.  Specifically, the A-B $(l_{Mo-Se})$ bond length in Fig.~\ref{fig:mos2_mosse}(c) is equal to $2.53~\text{\AA}$, while the A-C  bond length $(l_{Mo-S})$ is $2.416~ \text{\AA}$. The bending deformation of MoSSe to a curvature of $K_{eff}$ = 0.05 nm$^{-1}$ stretches $l_{Mo-Se} (l_2)$ from 2.53 to $2.72~\text{\AA}$ and shrinks $l_{Mo-S} (l_1)$ from 2.416 to $2.20~\text{\AA}$. The initial bond length asymmetry in MoSSe is further increased due to the given deformation and helps to produce larger dipole moments $p^z$, compared to MoS$_2$.  The dipole moment $p^{z}$ is also related to the polarizability of the atomic system, where the polarizability of MoSSe $(\alpha_{\text{MoSSe}})$ is 1.09 times larger than that of MoS$_{2}$ $(\alpha_{\text{MoS}_2})$ as shown in Table.~\ref{table:atomic_polar}, which shows the important effect of structural asymmetry for the Janus TMDCs. 

In addition, the initial bond length asymmetry in unbent MoSSe induces an intrinsic electric field, which is not present in MoS$_2$, as shown in recent DFT simulations \cite{Jin2018}. That work also reported the non-overlapping of out-of-plane wavefunctions for electrons and holes due to this electric field, which implies a weak bonding between the electron-hole pair, and which reduces the bandgap by pushing the $d$ orbitals of the metal atom closer to the Fermi level \cite{Er2018}.  The $d$ orbital shifting may enhance the charge transfer process through $\pi-\sigma$ coupling, which represents an easy transfer of charges from the S or Se atom to the Mo atom.  This phenomenon is reflected in our CD model as the calculated charge on Mo (atom A in Fig.~\ref{fig:mos2_mosse}(d)) in MoSSe is $0.733 e$, which is significantly larger than for Mo in MoS$_2$ (atom A in Fig.~\ref{fig:mos2_mosse}(b)), which is $0.282 e$, and as such the charge acquired by the Mo atom in MoSSe is $2.6$ times higher than in MoS$_2$.
 
For a curvature of $K_{eff}$=0.05 nm$^{-1}$, the charge transfer-induced change of $E^{q-z}$ in MoSSe is 117.6 $\text{V/\AA}$, which is exactly $2.6$ times higher than the field induced in MoS$_2$, and which reflects stronger $\pi-\sigma$ coupling in MoSSe.  Furthermore, the value of $E^{p-z}$ is higher in MoSSe (38.4 $\text{V/\AA}$) than in MoS$_2$ (7.4 $\text{V/\AA}$), which represents a stronger dipole interaction ($\sigma-\sigma$ coupling) in MoSSe than MoS$_{2}$.  Recent studies on the electronic properties of strained Janus TMDCs \cite{Huang2018} suggest an increased coupling between the $p$ orbitals of S/Se atoms with the in-plane bonding orbitals $d_{x^2-y^2}$ and $d_{xy}$ of Mo atom as a function of changes in bond angle. In MoSSe, the angle B-A-C in Fig. \ref{fig:mos2_mosse} varies from $81.1^\circ$ to $74.2^\circ$ between the initial and deformed states ($K_{eff}$ = 0.05 nm$^{-1}$), while for MoS$_2$, angle variations are  $81.93^\circ$ to $78.66^\circ$ for the initial and deformed states. The reduction in bond angles between MoSSe and MoS$_2$ may increase the contribution of $E^{p-z}$ in MoSSe.

Overall, $E^{q-z}$ is higher than $E^{p-z}$ in MoSSe, which represents that $\pi-\sigma$ coupling is dominant over $\sigma-\sigma$ coupling.  DFT simulations have shown that coupling between $d_{z^2}$ and $p$ orbitals ($\pi-\sigma$ coupling) is stronger than the coupling between $p$ and  $d_{x^2-y^2}$ and $d_{xy}$ ($\sigma-\sigma$ coupling) orbitals \cite{Huang2018,Er2018} in Janus TMDCs. The total electric field increment $\Delta(E^{p-z}+E^{q-z})$ in MoSSe is 156.07 $\text{V/\AA}$ at a curvature of $K_{eff}$ = 0.05 nm$^{-1}$, which is $2.94$ times higher than in MoS$_2$ at the same curvature. The increased polarizability and electric field increases the total dipole moment of MoSSe to $3.24$ times higher than in MoS$_2$.  The flexoelectric constant for MoSSe $(\mu_{\text{MoSSe}})$ is found to be $0.117$ nC/m, which is $3.6$ times higher than flexoelectric coefficient of MoS$_2$, which is consistent with the larger polarizability of MoSSe. The polarizability is directly related to the dielectric constant ($\varepsilon$) of the material \cite{Pan2016}. For instance, $\alpha$ of MoSSe and MoS$_2$ are 13.45 and 12.35 $\text{\AA}^3$, respectively as shown in Table~\ref{table:atomic_polar}.  The DFT calculated values for $\varepsilon$ for these materials are 8.67 and 8.05 \cite{Jin2018}.  Thus, the increased flexoelectric constant for MoSSe over MoS$_2$ is in agreement with the fact that felxeoelectric effect scales with the material dielectric constant \cite{Yudin2013,Zubko2013}.

As previously shown in Table \ref{table:piezo-flexo}, the in-plane piezoelectric coefficient for MoSSe is comparable with that of MoS$_{2}$.  However, the out-of-plane flexoelectric coefficient is $3.6$ times higher than MoS$_{2}$ (Table \ref{table:piezo-flexo}), which implies that there are relative benefits to bending flexoelectricity as compared to in-plane piezoelectricity when comparing Janus TMDCs to standard TMDCs. In order to understand this further, we compare the change in total electric field and asymmetry in bond length between in-plane tensile deformation and out-of-plane bending deformation for MoSSe for the same strain energy density, which is defined as the sum of atomic stress times the atomic strain over the volume of the deformed system. The change in $E^{q-z}$ and $E^{p-z}$ for bending deformation were previously noted as 117.6 and 38.4 $\text{V/\AA}$ for a curvature of $K_{eff}$ = 0.05 nm$^{-1}$, while for tensile deformation the values are 11.9 and 18.7 $\text{V/\AA}$, respectively. This shows that the induced electric fields are higher in bending than in tension, which generates high dipole moments. It is also observed that the difference in bond length ($l_2-l_1$) for bending is higher than in tension, for the same unit cell compared in tension and bending.  This reflects the larger bond length asymmetry that is induced in bending, which supports the charge transfer and enhanced $\pi-\sigma$ coupling based electric fields. 

To further compare the resulting electromechanical coupling between standard TMDCs and Janus TMDCs, we have computed the electrical energy density for atomic configurations at the same strain energy density for MoSSe and MoS$_{2}$. The electrical energy density is defined as the sum of the dot product between induced polarization and electric field over the volume. The atomic configuration of MoSSe at $K_{eff}$=0.05 nm$^{-1}$ gives an electrical energy density of $2.6~\times10^9$ J/m$^3$ and a strain energy density of $1.13~\times10^9$ J/m$^3$. By selecting an atomic configuration for MoS$_2$ under bending deformation with the same strain energy density results in an electrical energy density that is about 39\% of the MoSSe electrical energy density.  The electrical energy density under tensile deformation of MoSSe is about 84\% of the electrical energy density for MoSSe under bending.  These results show that the higher values of electric fields and large dipole moments under bending deformation point to the advantage of bending as compared to stretching in generating strong electromechanical coupling in Janus TMDCs.

\subsection{Flexoelectric effect among Janus TMDCs}
\label{sec:flexomxy}

From the previous section, it is clear that the asymmetry in bond lengths between layers of MXY induce large dipole moments through the increase in induced electric fields.  Table~\ref{table:piezo-flexo} lists the out-of-plane bending flexoelectric coefficients of the Janus TMDCs along with the initial bond length difference $l_2-l_1$ in Fig. \ref{fig:mos2_mosse}, from which a positive correlation is identified for both the MoXY and WXY Janus TMDCs.  We note that the out-of-plane piezoelectric coefficients also show a similar dependence on bond length asymmetry \cite{dong2017large}.  We mechanistically examine this correlation further using the electric fields due to charge-dipole ($E^{q-z}$) and dipole-dipole ($E^{p-z}$) interactions from the CD model.

Under the prescribed bending scheme, the increase in the total electric field $E^{q-z} + E^{p-z}$ increases with strain gradient $K_{eff}$ for every Janus TMDC. However, the relative contribution from $E^{q-z}$ or $E^{p-z}$ to  $E^{q-z}$+$E^{p-z}$ varies between elements of the Janus TMDCs group. In the MoXY group for a strain gradient of $K_{eff}$ = 0.1 nm$^{-1}$, the contribution from the charge-dipole interaction induced electric field $E^{q-z}$ in MoS$_{2}$, MoSSe, MoSeTe, and MoSTe to the increase in the total electric field ($E^{q-z}$+$E^{p-z}$) is 84.76\%, 75.48\%, 70.33\% and 56.29\%, respectively. The contribution from the dipole-dipole interaction induced electric field $E^{p-z}$ to the increase in the total electric field ($E^{q-z}$+$E^{p-z}$) is then 15.24\%, 24.52\%, 29.67\% and 43.71\% in MoS$_{2}$, MoSSe, MoSeTe, and MoSTe, respectively. 

The bending deformation further develops a coupling among the induced dipole moments via the $\sigma-\sigma$ interactions, which raises the contribution of $E^{p-z}$. For example in MoSTe, at $K_{eff}$ = 0.01 nm$^{-1}$, $E^{q-z}$ and $E^{p-z}$ are 56.82 ($\text{V/\AA}$) and 33.99 ($\text{V/\AA}$), respectively. When $K_{eff}$ is increased to 0.05 nm$^{-1}$, these values increase to 280.5 ($\text{V/\AA}$) and 216.99 ($\text{V/\AA}$).  The enhanced dipole-dipole interaction $E^{p-z}$ is due to the reduction in bond angle via the increased bond length asymmetry due to bending. The decrease in bond angle (B-A-C in Fig.~\ref{fig:mos2_mosse}) between initial and deformed states ($K_{eff}$ = 0.05 nm$^{-1}$) is $6.9^\circ$ for MoSSe and $7.4^\circ$ for MoSTe. The difference in $E^{p-z}$ for atom A in MoSSe is $0.38$ $\text{V/\AA}$ and for atom A in MoSTe is $0.53$ $\text{V/\AA}$ (not shown in Fig.~\ref{fig:mos2_mosse}). This further confirms that the increased reduction in the bond angle helps in increasing the dipolar interactions ($\sigma-\sigma$ coupling), which are measured in the form of $E^{p-z}$. Note that atoms A, B, and C are selected at exactly the same unit cell locations in MoSSe and MoSTe. The cumulative effect of bond angle reduction within the unit cell and across the unit cell in the given atomic system makes the contribution of $E^{p-z}$ significant. The total effect of increased electric fields and polarizability helps in inducing high polarization and thus high flexoelectric coefficient for MoSTe over other Janus TMDCs.  A similar trend is found in the WXY group in Table~\ref{table:piezo-flexo}, where high charge transfer and higher bond angle reduction in WSTe compared to other elements in WXY group. This makes the flexoelectric coefficient of WSTe higher than WSSe and WSeTe.

Table.~\ref{table:piezo-flexo} also shows that the MoXY group exhibits higher out-of-plane bending flexoelectric coefficients than the WXY group.  For example, MoSTe and WSTe have the highest flexoelectric coefficients in their groups, where the flexoelectric coefficient of MoSTe is about 1.09 times larger than WSTe. At $K_{eff}$ = 0.05 nm$^{-1}$, the total electric field change in MoSTe is found to be 1.4 times larger than that of WSTe while the polarizability of MoSTe is 0.79 times that of WSTe. Though the polarizability of MoSTe is lower than WSTe, the larger induced electric fields results in the flexoelectric constant of MoSTe being slightly larger than that of WSTe.

\subsection{Bending against the initial spontaneous curvature}

Up to this point, we have studied the bending-induced flexoelectric response of MXY materials while applying the out-of-plane bending toward to their initial curling direction (towards S in the case of MoSTe). In this section, we apply bending deformation against the initial curling direction (towards Te in the case of MoSTe) to study the effect of bending direction on the induced polarization. We followed the same simulation procedure described in Sec \ref{sec:simulation_method} except for inserting a minus sign in Eq. (\ref{eq:out-bending}) to indicate that the applied deformation is opposite to the initial curling direction. 

\begin{figure}
	\centering
	\includegraphics[width=1.0\linewidth]{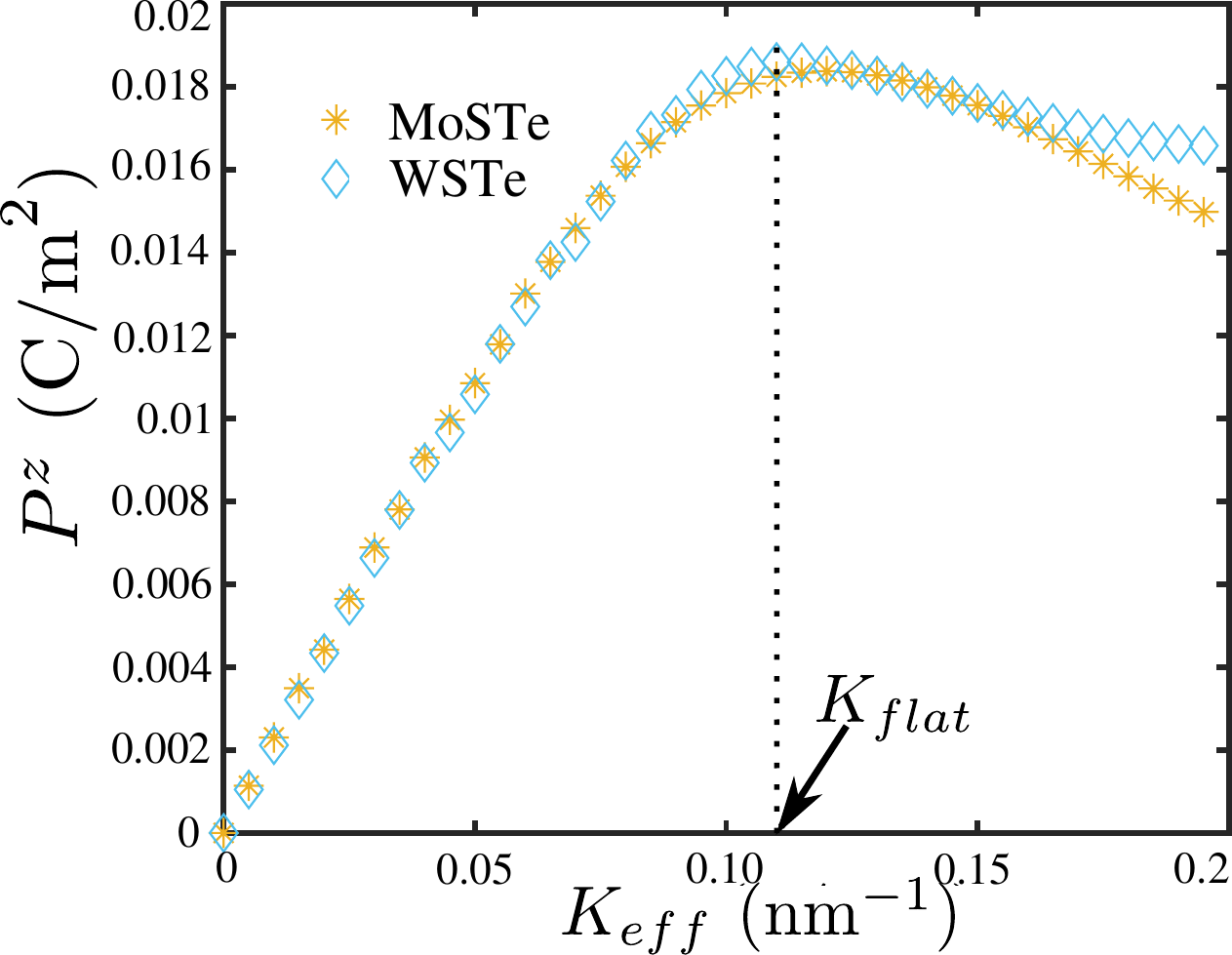}
	\caption{The variation of polarization $P^z$ with the effective strain gradient $K_{eff}$ for MoSTe and WSTe when bending against their initial spontaneous curvature.}
	\label{fig:pz_py_top_mxy}
\end{figure}

Fig.~\ref{fig:pz_py_top_mxy} shows the variation of $P^z$ with $K_{eff}$. The dashed line at $K_{eff}=K_{flat}$  represents the flattening of the MoSTe sheet from the initial spontaneously curved state.  The notable feature of Fig.~\ref{fig:pz_py_top_mxy} is that the total polarization $P^z$ increases up to $K_{flat}$ and decreases afterwards.  This is due to an increase in bond length between Mo and S atoms which reduces the charge transfer when $K_{eff}$ is larger than $K_{flat}$, which is seen in the form of decreasing $E^{q-z}$.  Specifically, the charge on an Mo atom at $K_{flat}$ and $1.4 \times K_{flat}$ are $0.55e$ and $0.43e$, respectively, which leads to a reduction in $E^{q-z}$.  A simultaneous decrease in $E^{p-z}$ is also observed due to an increase in the angle B-A-C, which increases from about $3.96^\circ$ from $K_{flat}$ to $1.4 \times K_{flat}$.  While the slope of the polarization $P^{z}$ variation with $K_{eff}$ is similar to the flexoelectric coefficient obtained in the case of bending towards the spontaneous curling direction, more energy is required to deform towards the Te layer in MoSTe as compared to the S layer in MoSTe.  As a result, bending towards the S layer in MoSTe is a better choice for energy conversion. A similar observation is found in the case of WSTe.  

\section{Conclusion} 
We have utilized classical molecular dynamics coupled with a charge-dipole model to study the bending flexoelectric response of Janus TMDCs.  We further employed a prescribed bending deformation that enabled us to directly calculate the flexoelectric constants while eliminating the piezoelectric contribution to the polarization.  In doing so, we found that Janus TMDCs have flexoelectric constants that are several times larger than traditional TMDCs such as MoS$_{2}$.  The mechanism underlying this was found to be bond length asymmetry for the Janus TMDCs between the M-X and M-Y atoms.  This bond length asymmetry was found to lead to stronger $\sigma-\sigma$ interactions with increasing initial asymmetry, along with stronger $\pi-\sigma$ interactions due to increased charge transfer, which combine to result in increased polarization for Janus TMDCs.

The present results also demonstrate the enhanced electromechanical coupling that results from out-of-plane bending as compared to in-plane stretching for Janus TMDCs.  Specifically, by comparing equivalent states of strain energy density, we found that the electrical energy density is higher for Janus TMDCs as compared to MoS$_{2}$, while the electrical energy density under tension is smaller than that of bending for Janus TMDCs.  These facts imply that there may be benefits in electromechanical energy conversion for Janus TMDCs by utilizing out-of-plane bending flexoelectricity as compared to in-plane piezoelectricity.

\section*{Acknowledgement}
\noindent The authors gratefully acknowledge the sponsorship from the ERC Starting Grant COTOFLEXI (No. 802205).


\bibliographystyle{apsrev4-1}
\bibliography{reference}

\end{document}